\documentclass[conference]{IEEEtran}

\usepackage{amsmath,amsfonts,amssymb,amsthm}
\usepackage{graphicx}
\usepackage{booktabs}
\usepackage{cite}
\usepackage{url}
\usepackage{subcaption}
\usepackage{multicol}
\usepackage{booktabs}
\usepackage{tabularx}
\usepackage{xcolor}
\usepackage{siunitx}

\DeclareMathOperator*{\argmax}{arg\,max}

\begin{document}

\title{FlightMagNav: An Open Dataset and Probabilistic Map Learning and Validation Framework for Outdoor Magnetic Field-Based Positioning}

\author{
\IEEEauthorblockN{Isaac Skog}
\IEEEauthorblockA{
KTH Royal Institute of Technology\\
Stockholm, Sweden\\
skog@kth.se
}
\and
\IEEEauthorblockN{Miguel Ramos Galrinho}
\IEEEauthorblockA{
FOI Swedish Defence Research Agency\\
Kista, Sweden\\
miguel.ramos.galrinho@foi.se
}
\and
\IEEEauthorblockN{Martin Gelin}
\IEEEauthorblockA{
FOI Swedish Defence Research Agency\\
Kista, Sweden\\
martin.gelin@foi.se
}
}

\maketitle

\begin{abstract}
Magnetic field-based positioning is a resilient positioning technology that requires no external infrastructure and is hard to jam at scale. To facilitate research on magnetic field-based positioning for aerial platforms operating close to the Earth's surface, an open dataset is presented with measurements collected using an unmanned aerial vehicle carrying two optically pumped magnetometers, a type of quantum magnetometer, and a global navigation satellite system-aided inertial navigation system. The dataset includes measurements for both magnetic-field map learning and validation. Along with the dataset, a probabilistic framework for magnetic-field map learning and validation is presented and used to illustrate how the dataset may be used. Finally, we outline research directions that may be explored using the dataset.
\end{abstract}


\section{Introduction}
Recent geopolitical developments have highlighted the need for resilient positioning technologies that can complement global navigation satellite systems (GNSS). One potential source of positioning information is the spatial anomalies in the Earth's magnetic field caused by ferromagnetic material in the Earth's crust. These magnetic anomalies are omnipresent and can be exploited for map-matching-based positioning in a way similar to using terrain height profiles in terrain navigation~\cite{Skog2026}. 

Recent advances in compact optically pumped magnetometers, a type of quantum magnetometer, have made it possible to equip small aerial platforms with highly sensitive and power-efficient magnetometers. Moreover, unlike classical flux-gate magnetometers, these magnetometers can measure magnetic-field magnitude directly, avoiding the need to reconstruct it from vector measurements. This reduces sensitivity to orientation and axis-misalignment errors, making it easier to reliably measure small magnetic-field anomalies~\cite{edelstein2007}. 

Currently, the development and evaluation of magnetic-field-based
positioning techniques for small aerial platforms is limited by the
availability of open datasets representative of small and slow-moving
platforms operating at low heights. Existing datasets for magnetic-field-based
positioning research, such as the DAF-MIT AIA airborne magnetic
navigation dataset~\cite{muradoglu2020}, have mainly been collected using
larger fast-moving platforms operating at greater heights. Low-height
magnetic-field datasets collected using unmanned aerial vehicles also
exist, see, e.g.,~\cite{phelps2022boulder,glen2021drone}. However, these
datasets have primarily been collected for geophysical surveying and for
studying platform disturbances, sensor noise, and magnetic compensation,
rather than for positioning research. Likewise, conventional magnetic-field
survey data, e.g.,~\cite{sgu2024magnetic}, are typically collected
with relatively coarse horizontal sampling and do not include data from
other navigation sensors, such as inertial sensors. Hence, available
survey data are typically not suitable for developing and evaluating
magnetic field-based positioning theories and methods. To that end, there
is a need for datasets of densely sampled magnetic-field and inertial
measurements collected on small, slow-moving platforms operating at low
heights.

This paper presents \emph{FlightMagNav}, an open dataset for outdoor magnetic field-based positioning and map learning. The dataset contains high-rate measurements from two optically pumped magnetometers and a GNSS-aided inertial navigation system collected during multiple unmanned aerial vehicle flights over an area of approximately 1~\unit{\kilo\meter\squared}. The dataset includes measurements from flights with lawnmower-pattern trajectories designed to learn magnetic anomaly maps, as well as flights with trajectories to validate the learned maps and study how the maps generalize to different operating heights.

In addition to the dataset, the paper presents a probabilistic framework for learning magnetic anomaly maps and for validating how well a learned map describes measurements from independent flights. The validation is performed by estimating the effective noise level using the marginal likelihood, i.e., the model evidence conditioned on the learned map. This provides a compact measure of how well the learned map generalizes to measurements from new flights and how the map's predictive accuracy changes with flight height.\\

\noindent\textbf{Reproducible research:} The code and data can be downloaded at \url{https://doi.org/10.71775/kth.ggkzd-jtk39}

\section{FlightMagNav Dataset}\label{sec:dataset}
Next, the data collection process for the \emph{FlightMagNav} dataset and the structure of the raw and parsed datasets are presented.

\begin{figure*}[t!]
	\centering
	\def\svgwidth{0.9\textwidth}
	\input{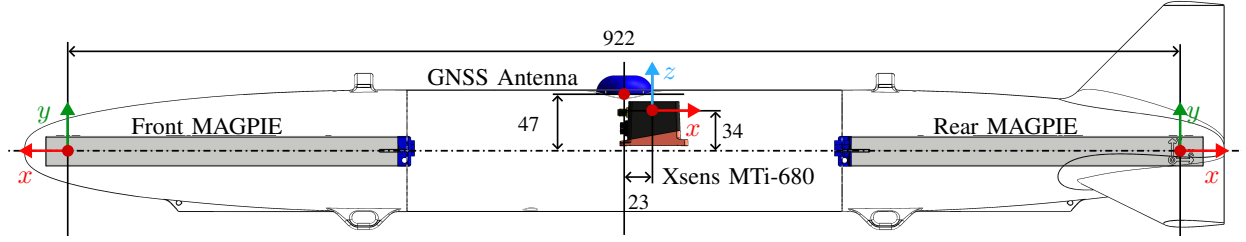}
	\caption{Placement of the sensors on platform with distances in \unit{\milli\meter}. Also shown are the coordinate axis of the sensors.}
	\label{fig:sensor placements}
\end{figure*}

\subsection{Sensor Platform}
A drawing of the sensor platform used during the data collection is shown in Fig.~\ref{fig:sensor placements}, along with the placement of the sensors. The platform carries two QuSpin MAGPIE sensor modules and one Xsens MTi-680 sensor module. Each QuSpin MAGPIE sensor module contains an optically pumped total-field magnetometer, a GNSS receiver, a low-cost inertial measurement unit, and an add-on coil for measuring the vector components of the magnetic field. The Xsens MTi-680 provides a GNSS-aided inertial navigation system (INS) solution. Tab.~\ref{tab:sensors} summarizes the sensors and the measurements they produce.

\subsection{Data Collection}
The data were collected outside Skövde, Sweden, over two days using a
quadcopter to carry the sensor platform shown in
Fig.~\ref{fig:sensor placements}. The sensor platform was suspended below
the quadcopter using two approximately \qty{8}{\meter} long ropes, allowing it to hang
freely during flight. Videos of the complete setup are included with
the published dataset. Temporal variations in the magnetic field were
monitored using a stationary QuSpin magnetometer located on the ground in the
vicinity of the flight area.

\begin{figure*}[t]
    \centering

    \begin{subfigure}[t]{0.48\textwidth}
        \centering
        \includegraphics[width=\linewidth]{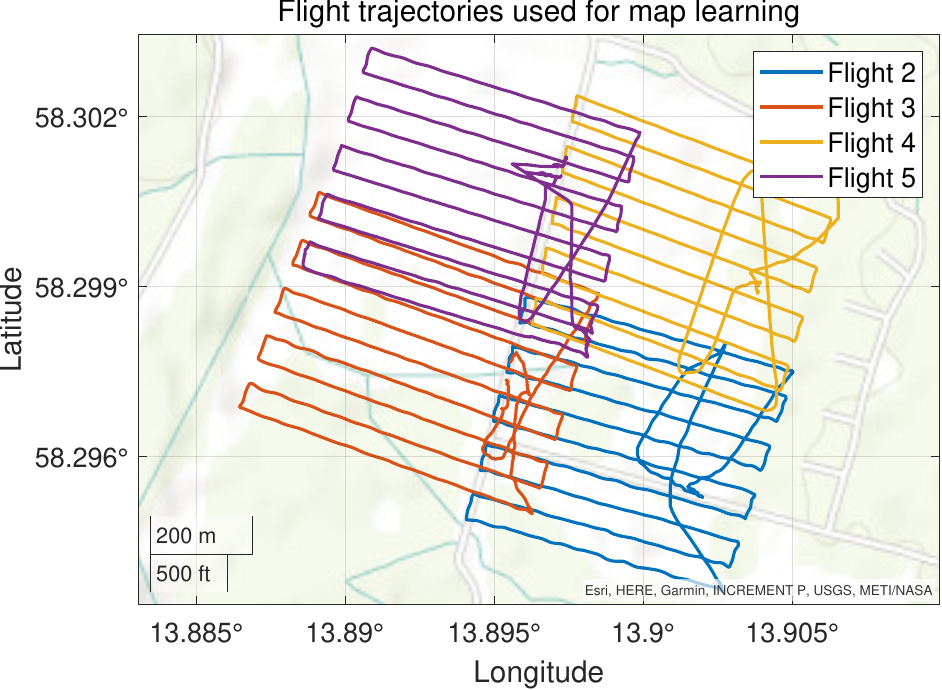}
        \caption{Map learning flight trajectories.}
        \label{fig:training_paths}
    \end{subfigure}
    \hfill
    \begin{subfigure}[t]{0.48\textwidth}
        \centering
        \includegraphics[width=\linewidth]{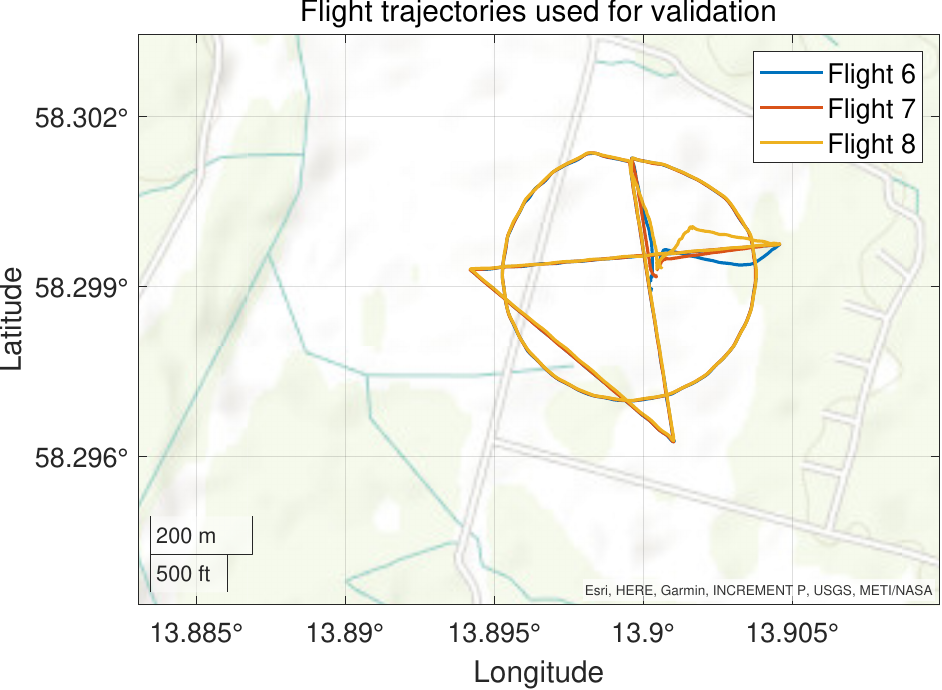}
        \caption{Validation flight trajectories.}
        \label{fig:validation_paths}
    \end{subfigure}

    \vspace{2mm}

    \begin{subfigure}[t]{0.48\textwidth}
        \centering
        \includegraphics[width=\linewidth]{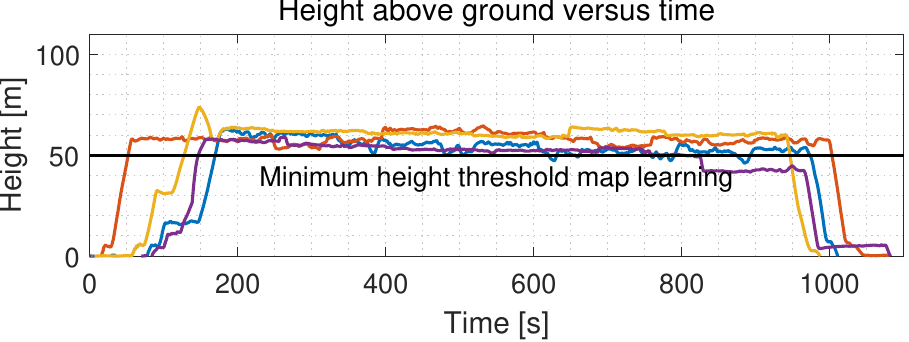}
        \caption{Map learning flight heights.}
        \label{fig:training_height}
    \end{subfigure}
    \hfill
    \begin{subfigure}[t]{0.48\textwidth}
        \centering
        \includegraphics[width=\linewidth]{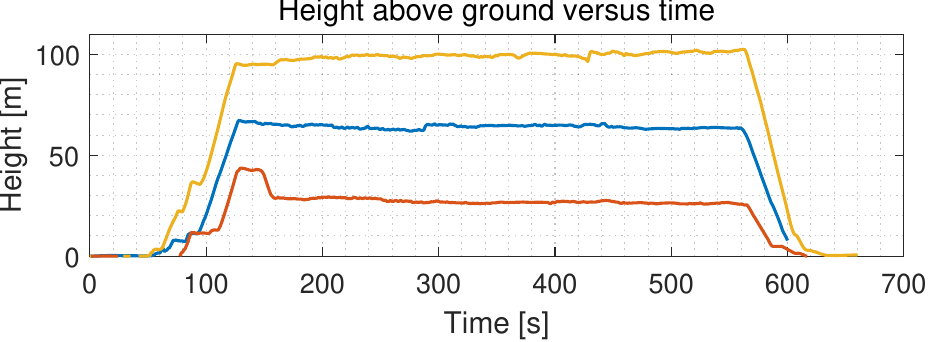}
        \caption{Validation flight heights.}
        \label{fig:validation_height}
    \end{subfigure}

    \caption{Flight trajectories and height profiles of the map learning and validation datasets. Note that the validation flights follow nearly the same horizontal trajectory but at different heights and remain entirely within the mapped area.}
    \label{fig:trajectories}
\end{figure*}

A total of eight flights were conducted. Flight~1 was a test flight used
to verify the functionality of the equipment. Flights~2--5 were designed for map learning and followed
lawnmower patterns covering different quadrants of an approximately
\qty{1}{\kilo\meter\squared} area. The separation between adjacent flight legs was
approximately \qty{50}{\meter}. The corresponding flight trajectories are shown in
Fig.~\ref{fig:training_paths}. During these flights, the sensor platform
was operated at a height of approximately \qty{55}{\meter} relative to the takeoff
point. The terrain over the flight area was approximately level, and the
height relative to the takeoff point closely corresponds to the
height above ground; see Fig.~\ref{fig:training_height}. 

Flights~6--8 were designed for validation of the learned map and for
investigating how well the map generalizes to flight heights different
from those used during map learning. The validation trajectories were all
well within the area covered by the map-learning flights, such that the
validation primarily tests generalization with respect to flight height
rather than horizontal extrapolation. The corresponding flight trajectories
and height profiles are shown in Fig.~\ref{fig:validation_paths} and
Fig.~\ref{fig:validation_height}, respectively.  During all flights, the speed
was approximately \qty{10}{\meter\per \second}. 

The local on-ground reference magnetometer did not provide usable data during
Flight~1, and the measurements from Flights~2 and~7 contain disturbances
caused by sensor movement and a passing vehicle. Therefore, magnetic-field
measurements from the Uppsala geomagnetic reference station, approximately \qty{270}{\kilo\meter} from Skövde, are also included in the dataset as an additional source of information about the
temporal variations in the magnetic field.
Despite this distance, during flights when the local reference sensor measurements were undisturbed, significant correlations can be observed between the temporal variations measured by the Uppsala reference station and those measured by the local sensor.

\begin{table*}[t]
	\caption{Sensors and sampling rates used during data collection.}
	\label{tab:sensors}
	\centering
	\small
	\begin{tabularx}{\textwidth}{
			@{}p{0.2\textwidth}
			p{0.2\textwidth}
			X
			p{0.08\textwidth}@{}
		}
		\toprule
		\textbf{Function} &
		\textbf{Sensor} &
		\textbf{Measurement} &
		\textbf{Sampling rate} \\
		\midrule
		
		Airborne magnetometers
		&
		$2\times$ QuSpin MAGPIE
		&
		Total magnetic field &
		\qty{125}{\hertz} \\
		
		(front and rear)  & with vector add-on &
		Vector magnetic field &
		\qty[parse-numbers=false]{125/3}{\hertz} \\
		& &
		Specific force and angular rate
		(from built-in ICM-20948 IMU) &
		\qty[parse-numbers=false]{125/2}{\hertz} \\
		
		& &
		Position
		(from built-in GNSS receiver) &
		\qty{1}{\hertz} \\
		
		\addlinespace
		
		Airborne GNSS-aided INS &
		Xsens MTi-680 &
		Position, velocity, attitude, specific force, and angular rate &
		\qty{100}{\hertz} \\
		
		\addlinespace
		
		Local on-ground reference magnetometer &
		QuSpin QTFM Gen-2 &
		Total magnetic field &
		\qty{125}{\hertz} \\
		
		\bottomrule
	\end{tabularx}
\end{table*}

\subsection{Dataset Organization}
The published dataset contains both the raw data as acquired by the sensors
and a MATLAB data file, \texttt{FlightMagNav.mat}, containing parsed and
synchronized data. The raw data are included to enable independent
processing of the measurements, while the parsed data provide a convenient
starting point for map learning and localization experiments. The structure
and organization of the two data representations are briefly described
next.

\subsubsection{Raw Data}
The raw data are organized by acquisition rather than by individual
flight. Each acquisition folder contains the recordings from the airborne
sensors and, when available, the local on-ground reference magnetometer. Consequently,
an acquisition folder may contain measurements corresponding to more than
one flight. Data from the Uppsala geomagnetic reference station are stored
separately and shared between all flights collected on the same day. The
correspondence between flights, acquisition folders, and start and stop
times is given in the accompanying metadata file. The raw data files are
summarized in Tab.~\ref{tab:raw_data}.

\subsubsection{Parsed Data}
The parsed dataset is organized by flight, where each element of the
MATLAB structure contains the measurements and metadata corresponding to
one flight. The organization of the parsed data is summarized in
Tab.~\ref{tab:parsed_data}.

The GNSS-aided INS defines the common time reference for the parsed data and has a sample rate of \qty{100}{\hertz}. Unfortunately, the GNSS timestamps associated with the MAGPIE measurements were insufficiently accurate for direct synchronization because of an incorrect configuration during data collection. Therefore, the MAGPIE measurements were synchronized to the GNSS-aided INS by correlating angular-rate measurements from the inertial sensors in the different sensor modules. A synchronization accuracy of about \qty{50}{\milli \second} was achieved. 

The scalar total-field measurements from the front and rear MAGPIE sensors
were synchronized and resampled to the \qty{100}{\hertz} time grid of the GNSS-aided INS. Before resampling, the total-field measurements were low-pass filtered
with a cutoff frequency of \qty{45}{\hertz} to avoid aliasing of the \qty{50}{\hertz} power-line disturbance.

The vector magnetic-field measurements were also synchronized to the
GNSS-aided INS time reference. However, they were not upsampled to \qty{100}{\hertz}, and their original sampling rate of $125/3$~\unit{\hertz} was retained. Hence, the vector measurements are time-aligned with the GNSS-aided INS while retaining their original sampling rate. Further, the measurements from both the local reference magnetometer and the Uppsala
geomagnetic reference station were resampled and aligned with the
time grid of the GNSS-aided INS.

The vector magnetic-field measurements and the inertial measurements from
the different sensor modules are provided in their respective sensor
coordinate frames. No rotation into a common coordinate frame has been
applied in the parsed dataset.

\begin{table*}[t]
	\caption{Raw data files included in FlightMagNav. Files with the same
		name occur in different acquisition folders. An acquisition may contain
		data from more than one flight; in particular, \texttt{acquisition\_06}
		contains Flights~6 and~7.}
	\label{tab:raw_data}
	\centering
	\small
	\begin{tabularx}{\textwidth}{
			@{}p{0.22\textwidth}
			p{0.23\textwidth}
			p{0.12\textwidth}
			X@{}
		}
		\toprule
		\textbf{File name} &
		\textbf{Source} &
		\textbf{Flight coverage} &
		\textbf{Description} \\
		\midrule
		
		\texttt{FrontMag.txt} &
		Front magnetometer &
		Flights~1--8 &
		Raw output from the front QuSpin MAGPIE sensor module. \\
		
		\texttt{BackMag.txt} &
		Back magnetometer &
		Flights~1--8 &
		Raw output from the back QuSpin MAGPIE sensor module. \\
		
		\texttt{GNSSINS\_raw.mtb} &
		GNSS-aided INS &
		Flights~1--8 &
		Native Xsens recording from the MTi-680 sensor module. \\
		
		\texttt{GNSSINS.txt} &
		GNSS-aided INS &
		Flights~1--8 &
		Text export of the Xsens MTi-680 measurements used by the provided
		parsing software. \\
		
		\texttt{RefMag.txt} &
		Local on-ground reference magnetometer &
		Flights~2--8 &
		Measurements from the stationary QuSpin total-field magnetometer.
		The measurements are considered usable for Flights~3--6 and~8. \\
		
		\texttt{ups\_ref\_20260610.sec} &
		Uppsala reference station &
		Flights~1--3 &
		Magnetic-field measurements from the Uppsala geomagnetic reference
		station. \\
		
		\texttt{ups\_ref\_20260611.sec} &
		Uppsala reference station &
		Flights~4--8 &
		Magnetic-field measurements from the Uppsala geomagnetic reference
		station. \\
		
		\bottomrule
	\end{tabularx}
\end{table*}

\begin{table*}[t]
	\caption{Organization of the parsed FlightMagNav data. Each element
		\texttt{data(i)} contains the data and metadata corresponding to one
		flight.}
	\label{tab:parsed_data}
	\centering
	\small
	\begin{tabularx}{\textwidth}{
			@{}p{0.16\textwidth}
			p{0.20\textwidth}
			X@{}
		}
		\toprule
		\textbf{Structure} &
		\textbf{Field} &
		\textbf{Description} \\
		\midrule
		
		\texttt{data(i)} &
		\texttt{flight\_id} &
		Flight identifier. \\
		
		&
		\texttt{start\_time} &
		Start time of the flight in UTC. \\
		
		&
		\texttt{stop\_time} &
		Stop time of the flight in UTC. \\
		
		&
		\texttt{raw\_folder} &
		Raw acquisition folder containing the corresponding measurements. \\
		
		&
		\texttt{purpose} &
		Purpose of the flight: test, learning, or validation. \\
		
		&
		\texttt{GNSSaidedINS} &
		Parsed measurements and navigation solution from the Xsens MTi-680. \\
		
		&
		\texttt{front\_mag} &
		Parsed and synchronized measurements from the front MAGPIE sensor
		module. \\
		
		&
		\texttt{back\_mag} &
		Parsed and synchronized measurements from the back MAGPIE sensor
		module. \\
		
		&
		\texttt{ref\_mag} &
		Parsed measurements from the stationary local reference
		magnetometer. Empty for flights without usable reference measurements. \\
		
		&
		\texttt{UPS\_ref\_mag} &
		Parsed magnetic-field measurements from the geomagnetic
		reference station. \\
		
		\bottomrule
	\end{tabularx}
\end{table*}

\section{Map Learning}\label{sec:learning}
Next, a probabilistic framework for learning a stochastic magnetic-field
anomaly map $\mathcal{M}$ while jointly accounting for sensor calibration
parameters is presented.

\subsection{Measurement Preprocessing}\label{sec:preprocessing}
To enable efficient and reliable map learning, the collected high-rate
measurements are preprocessed to compensate for known sensing artifacts and
disturbances, and downsampled to a suitable rate. Considering the collected
measurements and the subsequent map-learning process, the following
preprocessing steps are applied: (i) zero-phase anti-alias filtering and
downsampling of the measurements to suppress power-line disturbances and
reduce the number of measurement points; (ii) removal of measurement points
corresponding to heights below the height at which
the map is to be constructed; and (iii) subtraction of the nominal magnetic
field magnitude $m_{\text{\tiny nom}}$ from the measurements.

\subsection{Map Representation}
A variety of mathematical models can be used to represent a magnetic
map $\mathcal{M}(r)$. Many of these models can be expressed in terms of
a basis-function expansion of the form~\cite{Skog2026}
\begin{subequations}
\begin{equation}\label{eq:basisFunctions}
	\mathcal{M}(r)
	=
	\Phi^\top(r)w
	=
	\sum_{l=1}^{L}\phi_l(r)w_l,
\end{equation}
where
\begin{equation}
	\Phi(r)
	=
	\begin{bmatrix}
		\phi_1(r)\\
		\vdots\\
		\phi_L(r)
	\end{bmatrix}\quad\text{and}\quad
	w=
	\begin{bmatrix}
		w_1\\
		\vdots\\
		w_L
	\end{bmatrix}.
\end{equation}
Here, $\phi_l(r)$ and $w_l$ denote the $l$th basis function and its
corresponding weight, respectively. Further, $L$ denotes the number of basis
functions used to represent the map. In this work, radial basis
functions are used, i.e., 
\begin{equation}\label{eq:basis function}
	\phi_l(r)
	=
	\exp\left(
	-\frac{\|r-c_l\|^2}{2\ell^2}
	\right),
\end{equation}
\end{subequations}
where $c_l$ denotes the center of the $l$th basis function and $\ell$
denotes the length scale. With normally distributed weights, this basis-function representation defines a finite-rank approximation of a Gaussian process model. A sparse model, enabling efficient large-scale mapping at the cost of not being able to represent long-distance correlations in the map, can be constructed using compactly
supported radial basis functions~\cite{KullbergA2021}. 

\subsection{Measurement Model}
After preprocessing, the joint measurements $y^{(i)}_t$ from the front and back
magnetometers at time instant $t$ during flight $i\in\{1,\ldots,\mathrm{I}\}$ are assumed to be
independently distributed according to
\begin{equation}\label{eq:meas model}
	y^{(i)}_t\mid w,\xi^{(i)}
	\sim
	\mathcal{N}\left(
	A(x^{(i)}_t)w+B(x^{(i)}_t)\xi^{(i)},
	\sigma^2_{n^{(i)}}I_2
	\right).
\end{equation}
Here, $\xi^{(i)}$ and $\sigma^2_{n^{(i)}}$ denote the unknown calibration
parameters and the effective noise variance of the measurements collected during flight $i$, respectively.
Further, $A(x)$ and $B(x)$ denote the map and sensor-calibration
regression matrices, respectively.

The map regression matrix is given by
\begin{subequations}
\begin{equation}
	A(x)
	=
	\begin{bmatrix}
		\Phi^\top(r_f)\\
		\Phi^\top(r_b)
	\end{bmatrix},
\end{equation}
where $r_f$ and $r_b$ denote the positions of the front and back
magnetometers, respectively. These positions are given by
\begin{equation}
	r_j
	=
	r
	+
	R^{n}_{b}(q)d_j,
	\qquad j\in\{f,b\}.
\end{equation}
\end{subequations}
Here, $r$ and $q$ denote the position and orientation of the platform,
respectively. Further, $d_j$ denotes the position of the $j$th magnetometer in
the body frame (see Fig.~\ref{fig:sensor placements}), and
$R^{n}_{b}(q)$ denotes the rotation matrix transforming a vector from
the body frame to the navigation frame.

The sensor-calibration regression matrix is given by
\begin{subequations}
\begin{equation}
	B(x)
	=
	\begin{bmatrix}
		\psi(q)^\top & 1 & 0_{1\times3} & 0\\
		0_{1\times3} & 0 & \psi(q)^\top & 1
	\end{bmatrix},
\end{equation}
where
\begin{equation}
	\psi(q)
	=
	R^b_n(q)\frac{m_{0}}{\|m_{0}\|}.
\end{equation}
\end{subequations}
Here, $m_{0}$ denotes the nominal Earth magnetic-field
vector in the navigation frame. To also compensate for
platform-induced disturbances, the calibration model may be replaced
by a Tolles--Lawson model or extensions thereof~\cite{hagerA2026}.

For the subsequent inference process, define the parameter vector and the
per-flight measurement vector as
\begin{equation}
	\theta
	=
	\begin{bmatrix}
		w\\
		\xi^{(1)}\\
		\vdots\\
		\xi^{(\mathrm{I})}
	\end{bmatrix}
	\quad\text{and}\quad
	y^{(i)}
	=
	\begin{bmatrix}
		y^{(i)}_1\\
		\vdots\\
		y^{(i)}_{T^{(i)}}
	\end{bmatrix},
\end{equation}
respectively, where $T^{(i)}$ denotes the total number of sample instants. Similarly, define the per-flight regression matrices 
\begin{equation}
	A^{(i)}
	=
	\begin{bmatrix}
		A(x^{(i)}_1)\\
		\vdots\\
		A(x^{(i)}_{T^{(i)}})
	\end{bmatrix}
	\quad\text{and}\quad
	B^{(i)}
	=
	\begin{bmatrix}
		B(x^{(i)}_1)\\
		\vdots\\
		B(x^{(i)}_{T^{(i)}})
	\end{bmatrix}.
\end{equation}
The full regression matrix for flight $i$ can then be written as
\begin{equation}
	H^{(i)}
	=
	\begin{bmatrix}
		A^{(i)}
		&
		e_i^\top\otimes B^{(i)}
	\end{bmatrix}.
\end{equation}
Here, $e_i$ denotes the $i$th standard basis vector of dimension $\mathrm{I}$.

Finally, define the measurement vector for all flights in the training
dataset as
\begin{equation}
	\mathrm{y}
	=
	\begin{bmatrix}
		(y^{(1)})^\top & \cdots & (y^{(\mathrm{I})})^\top
	\end{bmatrix}^\top.
\end{equation}
The measurement model for all measurements can then be written as
\begin{subequations}
\begin{equation}\label{eq:stacked meas model}
	\mathrm{y}\mid\theta
	\sim
	\mathcal{N}(H\theta,R),
\end{equation}
where
\begin{equation}
	H
	=
	\begin{bmatrix}
		H^{(1)}\\
		\vdots\\
		H^{(\mathrm{I})}
	\end{bmatrix}
	\quad\text{and}\quad 
	R
	=
	\bigoplus_{i=1}^{\mathrm{I}}
	\sigma^2_{n^{(i)}}I_{2T^{(i)}}.
\end{equation}
Here, $\oplus$ denotes the matrix direct sum.
\end{subequations}

\subsection{Parameter Inference}
Assume independent zero-mean normal-distributed priors for the map weights $w$ and
calibration parameters $\xi$. The joint prior can then be written as
\begin{subequations}
\begin{equation}\label{eq:prior}
	\theta\sim\mathcal{N}(0,P_{\theta_0}),
\end{equation}
where
\begin{equation}
	P_{\theta_0}
	=
	\sigma_w^2 I_L
	\oplus
	\sigma^2_{\xi}I_{8I}.
\end{equation}
\end{subequations}
Here, $\sigma_w^2$ and $\sigma^2_\xi$ denote the prior variances
of the map weights and calibration parameters, respectively.

Since the measurement model \eqref{eq:stacked meas model} is linear in the parameters and follows a normal distribution, and the prior
\eqref{eq:prior} is also normally distributed, the posterior distribution is also
normally distributed. That is,
\begin{subequations}
	\begin{equation}\label{eq:posterior}
		\theta\mid\mathrm{y}
		\sim
		\mathcal{N}(\mu_\theta,P_\theta),
	\end{equation}
	where
	\begin{align}
		\mu_\theta
		&=
		P_{\theta y}P_{yy}^{-1}\mathrm{y},\\
		P_\theta
		&=
		P_{\theta_0}
		-
		P_{\theta y}P_{yy}^{-1}P_{\theta y}^\top,
	\end{align}
	and
	\begin{align}
		P_{yy}
		&=
		HP_{\theta_0}H^\top+R,\\
		P_{\theta y}
		&=
		P_{\theta_0}H^\top.
	\end{align}
\end{subequations}
The inference can also be performed recursively using standard
Bayesian recursive least-squares equations~\cite{Skog2026}, thereby avoiding the need
to form and store the potentially large matrices used in the batch solution.

\subsection{Posterior Map Distribution}
From the posterior distribution of $\theta$, the posterior distribution
of the map $\mathcal{M}$ at an arbitrary position $r$ can be calculated as
\begin{subequations}
\begin{equation}
	\mathcal{M}(r)\mid\mathrm{y}
	\sim
	\mathcal{N}\left(
	\mu_{\mathcal{M}}(r),
	\sigma^2_{\mathcal{M}}(r)
	\right),
\end{equation}
where
\begin{equation}
	\mu_{\mathcal{M}}(r)
	=
	\Phi^\top(r)\mu_w
\end{equation}
and
\begin{equation}
	\sigma^2_{\mathcal{M}}(r)
	=
	\Phi^\top(r)P_w\Phi(r).
\end{equation}
\end{subequations}
Here, $\mu_w$ and $P_w$ are the subvector and submatrix of $\mu_\theta$
and $P_\theta$, respectively, corresponding to the map weights $w$.

\subsection{Hyperparameter Learning}\label{sec:hyperparameter learning}
The hyperparameters $\eta$ of the measurement model and prior,
including the map length scale $\ell$, the effective noise variances
$\sigma^2_{n^{(i)}}$, $i=1,\ldots,\mathrm{I}$, and the prior variances $\sigma^2_w$ and $\sigma^2_\xi$, can be
learned by maximizing the model evidence. Specifically, the maximum
marginal likelihood estimate is given by~\cite{Theodoridis2020}
\begin{subequations}
\begin{equation}
	\hat{\eta}
	=
	\argmax_\eta
	\log p(\mathrm{y};\eta),
\end{equation}
where
\begin{equation}
	\begin{split}
		\log p(\mathrm{y};\eta)
		&=
		\log\int
		p(\mathrm{y}\mid\theta;\eta)
		p(\theta;\eta)
		\,d\theta\\
		&\propto
		-\mathrm{y}^\top P_{yy}^{-1}\mathrm{y}-\log|P_{yy}|.
	\end{split}
\end{equation}
\end{subequations}
\section{Map Validation}\label{sec:validation}
The model evidence can also be used to quantify how well the learned
map $\mathcal{M}$ describes magnetic-field measurements from flights
that were not included in the training data. This can be done by
estimating the effective noise variance of the measurements from a validation flight using
the marginal likelihood conditioned on the learned map. That is, the effective noise variance is estimated as
\begin{equation}\label{eq:effective noise variance}
	\hat{\sigma}^2_{n}
	=
	\argmax_{\sigma^2_{n}}
	\log
	p\left(
	y^{(j)}
	\mid
	\mathrm{y};
	\sigma^2_{n}
	\right),
\end{equation}
where $y^{(j)}$, $j\notin\{1,\ldots,\mathrm{I}\}$, denotes the measurements from
a validation flight. The marginal likelihood is obtained by
marginalizing over both the map weights and the calibration parameters, i.e.,  
\begin{subequations}
	\begin{equation}
		\begin{split}
			p\left(
			y^{(j)}
			\mid
			\mathrm{y};
			\sigma^2_{n}
			\right)
			&=
			\iint
			p\left(
			y^{(j)}
			\mid
			w,\xi^{(j)};
			\sigma^2_{n}
			\right)\\
			&\quad\cdot
			p(w\mid\mathrm{y})
			p(\xi^{(j)})
			\,dw\,d\xi^{(j)}\\
			&=
			\mathcal{N}\left(
			y^{(j)};
			\mu_{y^{(j)}},
			P_{y^{(j)}}
			\right),
		\end{split}
	\end{equation}
	where
	\begin{align}
		\mu_{y^{(j)}}
		&=
		A^{(j)}\mu_w,\\
		P_{y^{(j)}}
		&=
		A^{(j)}P_w(A^{(j)})^\top
		+
		\sigma^2_{\xi}B^{(j)}(B^{(j)})^\top\nonumber\\
		&+
		\sigma^2_{n}I_{2T^{(j)}}.
	\end{align}
\end{subequations}
Note that $\sigma^2_{\xi}$ is held fixed during the validation, with a value equal to that calculated during the map learning, and only
$\sigma^2_{n}$ is estimated from the validation data.
\section{Experimental Results}
The proposed map-learning and validation framework in Sec.~\ref{sec:learning} and Sec.~\ref{sec:validation} was applied to the data presented in Section~\ref{sec:dataset}. The results are as follows. 

\subsection{Map learning}
To learn a magnetic field anomaly map using the proposed framework, measurements from Flights 2 to 5 were preprocessed as described in Sec.~\ref{sec:preprocessing}. A low-pass cut-off frequency of \qty{0.45}{\hertz} was used, and the data were downsampled to \qty{1}{\hertz}. Furthermore, measurements collected at a height below \qty{50}{\meter} were removed. The model hyperparameters $\sigma^2_w$, $\sigma^2_\xi$, and $\sigma^2_{n^{(i)}}$, $i=2,\ldots,5$, were estimated using the method in Sec.~\ref{sec:hyperparameter learning}. The basis function centers were placed on a uniform grid with a spacing of \qty{50}{\meter}, and the length scale $\ell$ was set to \qty{60}{\meter}. The length scale was selected based on the rule of thumb that the shortest expected spatial wavelength is approximately equal to the flight height~\cite{canciani2016absolute}.

\begin{figure*}[tb!]
	\centering
	\includegraphics[width=\textwidth]{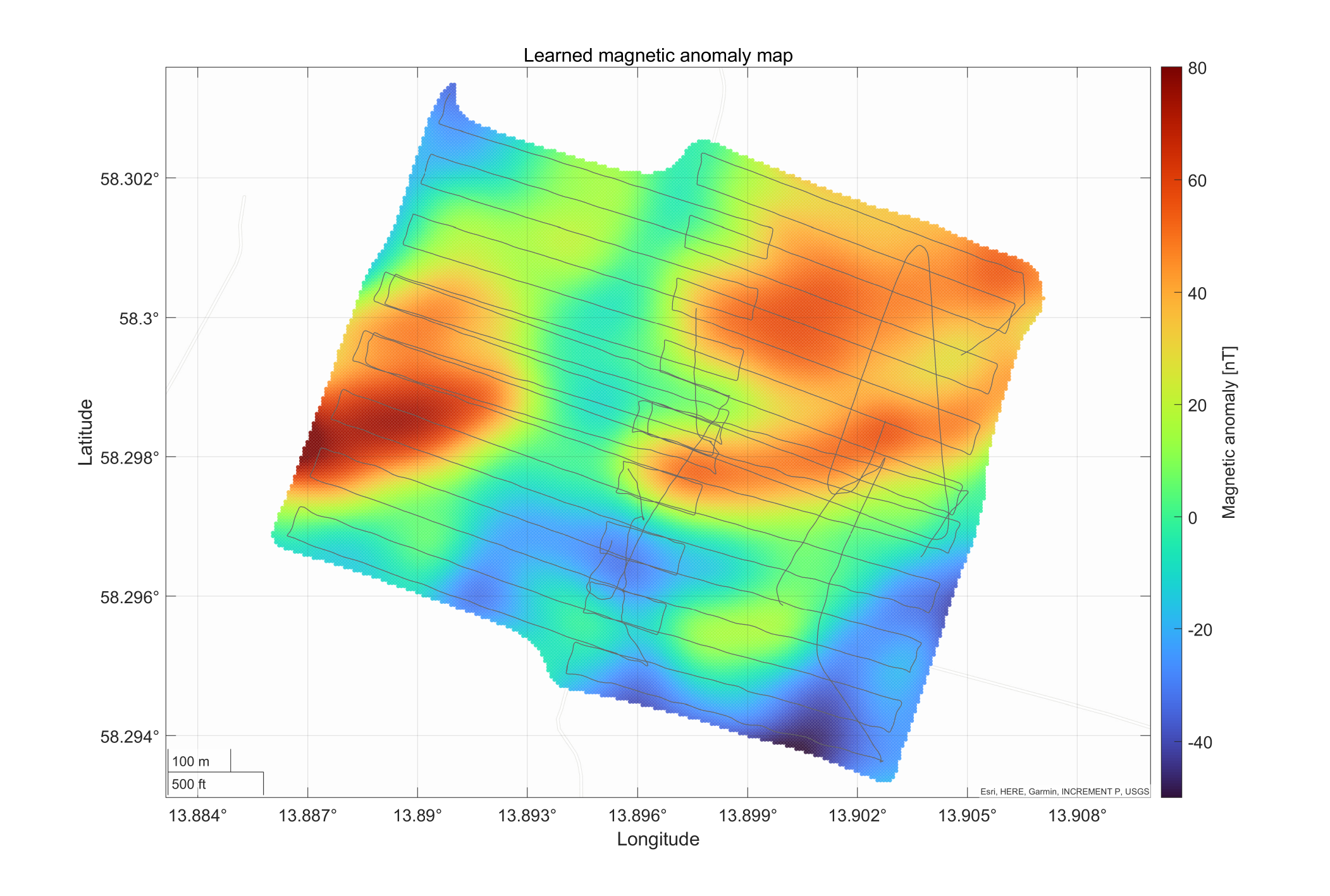}
	\caption{Expected value $\mu_\mathcal{M}(r)$ of the learned probabilistic magnetic field anomaly map $\mathcal{M}(r)$. Only areas where the posterior standard deviation $\sigma_\mathcal{M}(r)<$ \qty{3.5}{\nano\tesla} are shown. Also shown are the flight paths used in the training data collection.}
	\label{fig:map}
\end{figure*}

The learned map and the estimated hyperparameters are shown in Fig.~\ref{fig:map} and Tab.~\ref{tab:hyperparameters}, respectively. In the area covered by the flight paths, the inferred model indicates that the learned magnetic field anomaly map has an error with a standard deviation of less than \qty{3.5}{\nano\tesla}. Further, the estimated hyperparameters show that the effective noise standard deviation is approximately \qty{1}{\nano\tesla}, which is significantly higher than the specified sensor noise. At a \qty{0.5}{\hertz} bandwidth, the theoretical intrinsic sensor noise is \qty{2.1}{\pico\tesla}~\cite{skog2025underwater}. This indicates that the sensor noise itself is not the accuracy-limiting factor, but rather model errors and external field disturbances.  

\begin{table}[tb!]
	\caption{Estimated hyperparameters for the map learning.}
	\label{tab:hyperparameters}
	\centering
	\small
	\begin{tabular}{lc}
		\toprule
		\textbf{Hyperparameter} & \textbf{Estimate [nT]} \\
		\midrule
		$\hat{\sigma}_{n^{(2)}}$ & 0.6\\
		$\hat{\sigma}_{n^{(3)}}$ & 0.7 \\
		$\hat{\sigma}_{n^{(4)}}$ & 0.8\\
		$\hat{\sigma}_{n^{(5)}}$ & 1.3 \\
		$\hat{\sigma}_{\xi}$     & 9.6\\
		$\hat{\sigma}_{w}$       & 21.4 \\
		\bottomrule
	\end{tabular}
\end{table}

\subsection{Validation on Independent Flights}
To validate the quality of the learned magnetic field anomaly map and to illustrate the effect of using the map at heights other than those for which the map was created, the effective noise variances of the measurements from three validation flights were calculated using \eqref{eq:effective noise variance}. The flight paths and height profiles of the three validation flights are shown in Fig.~\ref{fig:validation_paths} and Fig.~\ref{fig:validation_height}, respectively. The flight paths are similar for all three flights, but heights differ. Excluding the take-off and landing phases, the flight height in Flight 6 was about the same as when the map was created, i.e., around \qty{60}{\meter}. During Flights 7 and 8, the heights were around 30 and 100 \unit{\metre}, respectively.

\begin{table}[t]
	\caption{Estimated effective noise for the validation flights.}
	\label{tab:effective_noise}
	\centering
	\small
	\begin{tabular}{ccc}
		\toprule
		\textbf{Flight} &
		\textbf{Average height} \textbf{[m]}  &
		\textbf{Effective noise} $\hat{\sigma}_n$ \textbf{[nT]} \\
		\midrule
		6 & 64 & 1.6 \\
		7 & 27 & 12.9 \\
		8 & 94 & 7.0 \\
		\bottomrule
	\end{tabular}
\end{table}

The estimated effective noise levels are shown in
Tab.~\ref{tab:effective_noise}. The results show that the predictive
accuracy of the learned map decreases when it is used at heights different
from that at which it was constructed. The degradation is particularly
pronounced when operating below the flight height used in the map-learning. 
This is expected because height acts as a spatial low-pass filter, with
frequency response
\[
H(\nu)\propto e^{-2\pi h\nu},
\]
where $h$ and $\nu$ denote height and spatial frequency, respectively.
Consequently, measurements collected below the map-learning height contain spatial-frequency components that are strongly attenuated at the height where the map-learning measurements were collected. Hence, the map does not represent these frequency components.         
\section{Summary and Outlook}
An open dataset for research on magnetic field-based positioning for
platforms operating close to the Earth's surface has been presented. The
dataset includes both raw and parsed, time-synchronized measurements from
two optically pumped magnetometers and a GNSS-aided INS. Along with the
dataset, a probabilistic framework for magnetic-field map learning and
validation has been presented and used to illustrate how the dataset may
be used. The validation framework quantifies the predictive accuracy of a
learned map in terms of an effective noise level, providing a scalar
measure for comparing maps.

The authors hope that the published data and the presented map-learning and
validation framework will support further research on magnetic field-based
positioning. Three research directions that may be studied using the
dataset are: (i) magnetic-field map matching and simultaneous localization
and mapping (SLAM) for unmanned aerial vehicles operating close to the
Earth's surface; (ii) magnetic-field map matching at heights below that
used for map learning, without relying on numerically ill-conditioned
downward continuation; and (iii) magnetic-field map learning and
positioning using combined total-field and vector-field measurements with different quality and sample rates.

\section*{Acknowledgment}
This work has been partially funded by Vinnova, Sweden’s Innovation Agency, and Swedish Armed Forces, project 2025-03855 \emph{Electromagnetic navigation for smaller unmanned underwater vehicles}. The authors would also like to thank \emph{Anton Åkesson} for constructing the housing for the measurement platform and \emph{Joakim Rydell} for serving as the pilot during data collection. 

\IEEEtriggeratref{7}
\bibliographystyle{IEEEtran}
\bibliography{IEEEfull,refs}

\end{document}